\documentstyle[pra,aps,twocolumn]{revtex}
\begin{document} 
\draft 
\title{Comment on ``Recurrences without closed orbits''}
\author{J\"org Main}
\address{Institut f\"ur Theoretische Physik I,
Ruhr-Universit\"at Bochum, D-44780 Bochum, Germany}
%
\date{\today}
\maketitle

\begin{abstract}
In a recent paper Robicheaux and Shaw [Phys. Rev. A {\bf 58}, 1043 (1998)]
calculate the recurrence spectra of atoms in electric fields with
non-vanishing angular momentum $\langle L_z\rangle\ne 0$.
Features are observed at scaled actions ``an order of magnitude shorter than
for any classical closed orbit of this system.''
We investigate the transition from zero to nonzero angular momentum and
demonstrate the existence of short closed orbits with $L_z\ne 0$.
The real and complex ``ghost'' orbits are created in bifurcations of the 
``uphill'' and ``downhill'' orbit along the electric field axis, and can
serve to interpret the observed features in the quantum recurrence spectra.

\end{abstract}


In Ref.\ \cite{Rob98} Robicheaux and Shaw calculate quantum photoabsorption
spectra of atoms in electric fields with nonzero magnetic quantum numbers,
$m\ne 0$ and observe recurrence peaks at short actions in the Fourier
transform recurrence spectra.
For spectra with magnetic quantum number $m=0$ these peaks can be
directly interpreted as the recurrences of the ``uphill'' and ``downhill''
orbit along the electric field axis.
For nonzero angular momenta the authors argue that ``these two orbits are
not possible, because $L_z$ is conserved and there is a repulsive
$L_z^2/(x^2+y^2)$ term in the potential.''
The observed features at short actions are therefore interpreted as 
``recurrences without closed orbits.''
It is the purpose of this comment to demonstrate that the uphill and downhill 
orbit do not disappear to nowhere at the transition from zero to nonzero 
angular momentum $L_z$, but, by contrast, closed orbits with approximately 
the same short action still exist for $L_z\ne 0$.
As will be shown, the orbits along the $z$ axis undergo bifurcations and 
split into a real and a complex ``ghost'' orbit.
The importance of ghost orbits for the photoabsorption spectra of atoms in
a magnetic field has been discussed at length in \cite{Mai97}.

\begin{figure}[b]
\vspace{6.0cm}
\includegraphics{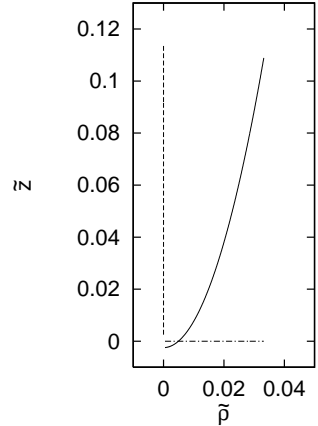}
\caption{\label{fig1} 
Closed orbits of the hydrogen atom in an electric field at scaled energy 
$\varepsilon=-4.0$ and angular momentum $\tilde L_z=0.014$ drawn in 
dimensionless scaled coordinates 
$(\tilde\rho=\rho F^{1/2}, \tilde z=zF^{1/2})$.
Solid line: Real orbit. 
Dashed and dash-dotted lines: Real and imaginary part of the complex 
``ghost'' orbit, respectively.
The orbits have bifurcated from the uphill orbit at vanishing angular 
momentum and have scaled action $\tilde S\approx (-2\varepsilon)^{-1/2}=0.35$.
}
\end{figure}

For the hydrogen atom in an electric field the Hamiltonian separates in 
semiparabolical coordinates, $\mu=\sqrt{r+z}$, $\nu=\sqrt{r-z}$, i.e.,
$H=H_\mu+H_\nu$ with
\begin{eqnarray}
\label{Hmu:eq}
 H_\mu &=& {1\over 2}p_\mu^2 - \varepsilon\mu^2
    + {\tilde L_z^2\over 2\mu^2} + {1\over 2}\mu^4 = 2Z_1 \; , \\
\label{Hnu:eq}
 H_\nu &=& {1\over 2}p_\nu^2 - \varepsilon\nu^2
    + {\tilde L_z^2\over 2\nu^2} - {1\over 2}\nu^4 = 2Z_2 \; ,
\end{eqnarray}
$Z_1+Z_2=1$, $\varepsilon=EF^{-1/2}$ the scaled energy, and 
$\tilde L_z=L_zF^{1/4}$ the scaled angular momentum.
$F$ is the electric field strength.
Obviously, for $L_z\ne 0$ the centrifugal barrier does not allow trajectories
to start exactly at the origin.
However, the shortest closed orbits can easily be derived from the conditions
on the time evolution
$p_\nu(\tau)=0$, $\nu(\tau)=\nu_0=\mbox{const}$ for the orbits bifurcating 
from the uphill orbit and 
$p_\mu(\tau)=0$, $\mu(\tau)=\mu_0=\mbox{const}$ for the orbits bifurcating 
from the downhill orbit.
In the following we discuss the bifurcation of the uphill orbit, for the
downhill orbit similar results are obtained at low energies 
$\varepsilon \ll -2.0$.
The effective potential $V(\nu)=-\varepsilon\nu^2+\tilde L_z^2/2\nu^2-\nu^4/2$
has a local minimum for energies $\varepsilon < -{3\over 2}\tilde L_z^{2/3}$.
The stationary $\nu$ motion is obtained from the condition of vanishing
derivative $dV(\nu)/d\nu=0$, yielding
\begin{equation}
 2\nu_0^6 + 2\varepsilon\nu_0^4 + \tilde L_z^2 = 0 \; .
\label{nu0:eq}
\end{equation}
Two approximate solutions of (\ref{nu0:eq}) at $\varepsilon \ll 0$ are
$\nu_0^2\approx \pm\tilde L_z/\sqrt{-2\varepsilon}$
approaching $\nu_0=0$ in the limit $\tilde L_z\to 0$.
The two solutions represent a real and a ghost orbit for $\nu_0$ real and
imaginary, respectively.
With given $\nu_0$ it is a straightforward task to calculate the constant 
of motion $Z_1=1-Z_2$ and to solve for $\mu(\tau)$ in Eq.\ \ref{Hmu:eq}.
The shapes of the closed orbits are presented in dimensionless scaled 
coordinates $(\tilde\rho=\rho F^{1/2}, \tilde z=zF^{1/2})$ in Fig.\ \ref{fig1}
for $\varepsilon=-4.0$ and scaled angular momentum $\tilde L_z=0.014$, which 
belongs to the magnetic quantum number $m=1$ at a value 
$\omega=2\pi\sqrt{\varepsilon/E}=450$.
This corresponds to the values chosen in Figure 1 of Ref.\ \cite{Rob98}.
The solid line is the real orbit, which is the uphill orbit distorted
by the repulsive centrifugal barrier.
An analogous orbit has been discovered and
%
\noindent
discussed for the diamagnetic 
Kepler system with non-vanishing angular momentum \cite{Nie91}.
The dashed and dash-dotted lines are the real and imaginary part of the
complex ghost orbit, respectively.
The real part of the ghost orbit is nearly identical with the uphill orbit
at vanishing angular momentum $\tilde L_z=0$.
The scaled action of both orbits is 
$\tilde S\approx 1/\sqrt{-2\varepsilon}=0.35$ in perfect agreement with the 
first recurrence peak in Figures 2 and 3 of Ref.\ \cite{Rob98}.

As mentioned above, for $L_z\ne 0$ the centrifugal barrier does not allow
trajectories to start exactly at the origin.
The nearest distance of closed orbits from the origin depends on the values
of the constants of motion $Z_1$ and $Z_2$ in Eqs.\ \ref{Hmu:eq} and 
\ref{Hnu:eq}.
It is negligible small for orbits with $Z_1\approx Z_2\approx 0.5$ and 
increases when $Z_1$ or $Z_2$ approaches the minimal allowed value.
For the real closed orbit in Fig.\ \ref{fig1} the nearest distance from the 
origin is $\tilde r_{\rm min}=r_{\rm min}F^{1/2}=0.0025$ in scaled units, 
which is about 13 Bohr radii at $\omega=2\pi\sqrt{\varepsilon/E}=450$.
This is slightly outside the classically allowed region of the initial 
state $|2p1\rangle$, however, it should be noted that a small change of the 
initial conditions results in approximately closed orbits where the distance 
to the origin at the start and return is reduced to about a few Bohr radii.
The real closed orbit in Fig.\ \ref{fig1} is more strongly excited in 
dipole transitions from initial states of larger size than the size of 
the hydrogenic $|2p1\rangle$ state as can clearly be seen in Figs.\ 5 and 6
of Ref.\ \cite{Rob98} for the excitation of the K and Cs atom, respectively.
The relatively large nearest distance of the closed orbit to the origin
suppresses effects of classical core scattering \cite{Hue95}, especially
for the K atom (see Fig.\ 5 in Ref.\ \cite{Rob98}), which might result in 
strong damping of the multiple repetitions for orbits with $L_z=0$ diving 
deeply into the ionic core.
For the Cs atom the ionic core has larger size and core scattering can be 
observed in Fig.\ 6 of Ref.\ \cite{Rob98}, but, in contrast to the 
interpretation given in \cite{Rob98}, orbits are scattered into the 
{\em truly existing} short closed orbit presented in Fig.\ \ref{fig1}.
However, it is still an outstanding task to reproduce quantitatively the 
amplitudes of recurrence peaks in the Fourier transform quantum spectra
of the hydrogen atom and non-hydrogenic atoms by application of closed 
orbit theory.
Robicheaux and Shaw are probably right that the theory may need to be 
generalized to account for the effects of orbits not starting at and 
returning back exactly to the origin.

In conclusion, we have investigated the bifurcation scenario of the shortest
closed orbits of the hydrogen atom in an electric field at the transition
from zero to non-vanishing angular momentum $L_z$, and revealed the existence
of short real and complex ghost orbits with $L_z\ne 0$.
They are born in bifurcations of the uphill and downhill orbit along the 
field axis and can serve to interpret features at short scaled actions in 
the quantum recurrence spectra calculated by Robicheaux and Shaw \cite{Rob98}.

\bigskip
This work was supported by the Deutsche Forschungsgemeinschaft.
I am grateful to G.\ Wunner for a critical reading of the manuscript.

\end{document}